\documentstyle[epsf]{amsart} 
\def\bbbone{{\mathchoice {1\mskip-4mu \text{l}} {1\mskip-4mu \text{l}}
{ 1\mskip-4.5mu \text{l}} { 1\mskip-5mu \text{l}}}}

\def\squarebox#1{\hbox to #1{\hfill\vbox to #1{\vfill}}} 
\newcommand{\stopthm}{\hfill\hfill\vbox{\hrule\hbox{\vrule\squarebox 
                 {.667em}\vrule}\hrule}\smallskip}

\newsymbol\circlearrowleft 1309
\newsymbol\restriction 1316 
\newcommand{\rest}{\!\!\restriction}

\theoremstyle{plain}

\newtheorem{lem}{Lemma}
   
\newtheorem{prop}{Proposition}
   
\theoremstyle{definition}

\numberwithin{equation}{section}

\title[Spacing between phase shifts]
{Spacing between phase shifts in a simple scattering problem}
\author[S. Zelditch and M. Zworski]{Steve Zelditch and Maciej Zworski}

\begin{document} 

\begin{abstract}
We prove a scattering theoretical version of the Berry-Tabor 
conjecture: for an almost every surface in a class of cylindrical 
surfaces of revolution, the large energy limit of
the pair correlation measure of the quantum phase shifts is Poisson, that
is, it is given by the uniform measure. 
\end{abstract}

\maketitle

\section{Introduction and statement of the result} 
The   Berry-Tabor conjecture \cite{BeTa} for quantum integrable systems
with discrete spectra  
 asserts   that the spacings between normalized eigenvalues of a 
quantum integrable system should exhibit Poisson statistics in the 
semi-classical limit.  In particular, when the eigenvalues are scaled
to have {\em unit mean level spacing}, the distribution
of their differences should be uniform.  This conjecture has been  
verified
numerically in many cases \cite{num} and has been rigorously proved in
an almost everywhere sense for flat 2-tori (Sarnak \cite{Sa}),  
flat 4-tori (Vanderkam \cite{V}), deterministically for almost all flat tori
(Eskin-Margulis-Mozes \cite{EMM}), 
and for certain
integrable quantum maps in one degree of freedom (Rudnick-Sarnak  
\cite{RuSa},
Zelditch \cite{Zel2}).  Smilansky \cite{Sm} has more recently posed an  
analogous
conjecture for scattering systems with continuous spectra.  Since
 the scattering matrix $S(E)$ 
at energy level $E$ is, at least heuristically, the quantization of the  
classical
scattering map, he argues that when the 
scattering map is integrable  the eigenvalues of $S(E)$ (
known as {\it phase shifts}) should exhibit Poisson statistics.  In  
particular, he proposed that the
pair correlation function of scaled phase shifts should be uniform for  
surfaces of revolution with a cylindrical end (see figure 1).  
The purpose of this paper is to prove (a somewhat modified form of)  
this
 conjecture for almost all surfaces in
an infinite dimensional family of (pairs of) such surfaces.  

To explain the modifications and  state our results, 
we need to introduce some notation.
The surfaces we consider are topological discs $X$ on which $S^1$
 acts freely  except
for a unique fixed point $m$.  
The metrics $g$ we consider are
invariant under the $S^1$ action and in geodesic polar 
coordinates centered at $m$ have
the form $g = dr^2 + a(r)^2 d\theta^2$ where $a(r)$ defines a short  
range 
{\it cylindrical
end} metric (see Sect.2).  
For technical reasons, we are only able to analyse the  phase
shifts at this time in the case where $g$ has a conic singularity at  
$m$.  
  
\begin{figure}[htb]
\hspace{0.0cm}
 \epsfbox{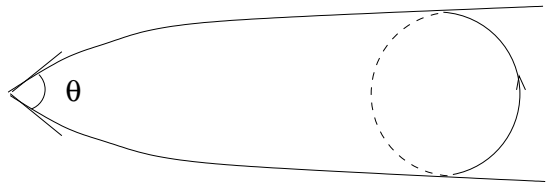}
\caption{A surface of revolution 
with a conic singularity and a cylindrical end.}
\label{Fig.1}
\end{figure}

To define the pair correlation measure, we recall  
that at energy $ \lambda^2 $ the scattering matrix for 
a surface of revolution with a cylindrical end is given by  
a diagonal $ ( 2 [ \lambda ] + 1 ) \times ( 2 [\lambda ] + 1 ) 
$ matrix with entries $ \exp ( 2 \pi i \delta_k ( \lambda ) ) $, 
 $ |k | \leq [\lambda ] $ -- see Sect.3 for a detailed presentation. 
The phase shifts are given by $ \delta_k ( \lambda ) $ and are 
well defined modulo $ {\Bbb Z} $. The parameter $ k $ corresponds 
to the angular momentum or in other words to the eigenvalues of the 
Laplacian on the cross-section (the circle in our case). When $  
|k| $ is close to $ \lambda $ we expect no scattering phenomena 
as the classical motion is close to the bounded motion along the 
cross-sections (see Fig.2). At the opposite extreme, when $  
|k | / |\lambda | $ is close to $ 0 $, the classical motion is  
along geodesics approaching the singularity on the surface. 

\begin{figure}[htb]
\hspace{0.0cm}
 \epsfbox{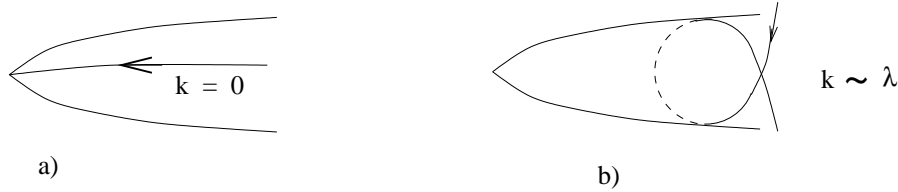}
\caption{a) Geodesics approaching the singularity; b) Geodesics close  
to 
the cross-sections.}
\label{Fig.0}
\end{figure}

Since the properties of the pair correlation measure are supposed to  
correspond to the properties of smooth classical motion it is  
natural, at least at this early stage, to delete the angular momenta 
corresponding to the neighbourhoods of the singularities. 
 
Based on this discussion we define for any $ \epsilon > 0 $ the 
following measure 
\begin{align}
\label{eq:pcm1} 
\begin{aligned}
\rho_\lambda^\epsilon \left(  
[ a , b ] \right) \stackrel{\text{def}}{=}   & 
\frac{1}{ ( 1 - 2 \epsilon ) } \frac{1 }{ 2 \lambda + 1 }  
\; \sharp \; \left\{  \; 
( l , m , k ) \; : \;  
l,m,k \in {\Bbb Z} \,, \ \epsilon < \left| {l}/{\lambda}  
\right|, \left| {m}/{\lambda}  
\right| < 1 - \epsilon  \,, \right. \\   & 
 \ \ \left. 
( 2 \lambda + 1 ) ( 1  -  
2 \epsilon )  ( \delta_l ( \lambda ) - \delta_m ( \lambda ) + k )   
\in [ a, b ] \; \right\} \,. 
\end{aligned}
\end{align}
In other words for $ f \in {\cal S}  ( {\Bbb R} )$, 
\begin{equation} 
\label{eq:pcm2} 
\int f ( x) \rho_\lambda^\epsilon ( d x) =  
\frac{1}{ ( 1 - 2 \epsilon ) } \frac{1 }{ 2 \lambda + 1 }  
\sum_{ k \in {\Bbb Z} } \sum_{ { m , l \in {\Bbb Z } }\atop{ 
\epsilon < |m/\lambda|, |l/ \lambda|< 1 - \epsilon }}  
f  \left( ( 1 - 2 \epsilon ) ( 1 + 2 \lambda ) (  
\delta_l ( \lambda ) - \delta _m ( \lambda) + k )\right) \,. 
\end{equation} 
Although rather cumbersome, this definition follows the standard 
procedure for defining pair correlation measures
 -- see 
the references listed above. 
 
Our main result concerns the (modified) pair correlation function for  
an 
infinite dimensional  set  ${\cal G}$
of 2-parameter families of surfaces of revolution  
\[ (X,g^{\alpha, \beta})\,, \ \ (\alpha, \beta) \in 
( \alpha_0 - \delta, \alpha_0 + \delta ) \times ( - \delta , 
\delta ) \subset {\Bbb R} ^2 \,,  \]
with cylindrical ends. The precise definition of ${\cal G}$ will be  
given in 
Proposition \ref{p:5} in Sect.5. 
The key property of the metrics is that
the leading parts of the
phase shifts, $\psi^{\alpha, \beta}$, of the 2-parameter families 
$(X, g^{\alpha, \beta})$ 
depend linearly on the parameters $(\alpha, \beta)$.  
This feature allows us to prove: 

\vspace{0.23cm} 
\noindent 
{\bf Theorem.} {\em 
Let $\{g_{\alpha, \beta}\} \in {\cal G}.$ Then for almost every
pair $(\alpha, \beta)$ (in the sense of Lesbesgue measure) and for any  
sequence
$ \{ 
\lambda_m \}_{m=0}^\infty $ satisfying 
\[ \sum_{m=0}^\infty  \frac{ \log^3 \lambda_m }{ \lambda_m } < \infty  
\,,
\] 
we have 
\begin{equation} 
\label{eq:thm} 
\lim_{m \rightarrow  
\infty}
\rho_{\lambda_m}^\epsilon ( f) = 
 \int f ( x) dx + f ( 0 ) \,, \ \ f \in {\cal S} ( {\Bbb R} ) \,, \ \ 
\epsilon > 0 \,. 
\end{equation} } 
 
\vspace{0.23cm} 

 This Theorem  proves the Berry-Tabor-Smilansky conjecture  
for phase shifts for our class of surfaces. The statement can 
only hold almost everywhere as we can produce one parameter families 
of surfaces for which the pair correlation measure is {\em not}  
uniform. The proof of the Theorem is based on Proposition \ref{p:4} below
which is a somewhat stronger and more precise result.

\medskip

\noindent{\sc Acknowledgments.} 
This note
 originated in a discussion between U.Smilansky and the first author at  
the Newton Institute
during the program on Quantum Chaos and Disordered Systems  in 1997  
following a talk on the
results of \cite{Zel1}.   Smilansky pointed out that the formal WKB  
formula for  phase shifts of a surface 
of revolution with cylindrical end   closely resembled the WKB formula  
for eigenvalues of an integrable quantum map on $S^2$ and proposed the  
problem of proving that the scaled phase shifts exhibit
Poisson statistics.  We would like thank U.Smilansky for bringing 
this to our attention.

The first author would also like to thank
the National Science Foundation for partial support under
the grant DMS-9703775. The second author would like to thank
the National Science and Engineering Research Council of
Canada for partial support. Both authors are grateful for the  
hospitality 
of the Erwin Schr\"odinger Institute where part of this work was done.

\section{Surfaces of revolution with cylindrical ends} 
We consider a class of incomplete two dimensional smooth manifolds  
denoted by $ X \setminus \{ m \} $, such that $ X$ is a topological 
completion of $ X \setminus \{ m \} $. We can then consider $ X $ as 
a manifold with a conic singularity. The manifold $ X \setminus \{ m \}$ 
is globally parametrized by $ (0, \infty ) \times {\Bbb S}^1 $ and we put 
on it metrics of revolution: 
\begin{equation} 
\label{eq:1.1} 
g = dr^2 + a ( r) ^2 d \theta^2 \,, \ \ r \in ( 0 , \infty ) \, ,  
\ \theta\in {\Bbb S}^1 \,.  
\end{equation} 
The metric is assumed to be a short range {\em cylindrical end} 
metric, that is, we require that  
\begin{equation} 
\label{eq:1.2} 
\left|\partial_r^k \left( a ( r)^2 - 1 \right) \right| \leq  
C_k r^{ - 2 - k } \,, \ \ r \longrightarrow \infty \,.  
\end{equation} 
At $ m $ we assume a conic structure: 
\begin{equation} 
\label{eq:1.2'} 
a( 0 ) = 0 \,, \ \ a'( 0 ) \neq 0 \,. 
\end{equation} 
We also make a convexity assumption by demanding that 
\begin{equation} 
\label{eq:1.3} 
a'( r) > 0 \,. 
\end{equation} 
The metric can be extended to a smooth metric on  $X $  
(endowed with a natural $ {\cal C}^\infty $ structure  
coming from polar coordinates, $ ( r , \theta ) $ ) if 
and only if 
\begin{equation} 
\label{eq:1.4} 
a' ( 0 ) = 1 \,, \ \ a^{2p} ( 0 ) = 0 \, , \ p \geq 0 \,, 
\end{equation} 
see for instance \cite{Bess}. We will not assume \eqref{eq:1.4} 
and consequently we allow {\em bullet like} surfaces shown in  
Fig.1. 
 
The classical dynamics is given by the Hamiltonian flow of the  
metric: 
\begin{equation} 
\label{eq:1.5} 
p = |\xi|^2_{g_x} = \rho ^2 + a(r)^{-2} t^2  \,, 
\end{equation} 
where we parametrized $ T^* ( X \setminus \{ m \} ) $ by  
 $ ( x , \xi ) = ( r , \theta ; \rho , t ) $, with $ \rho $ 
and  $ t $ dual to $ r $ and $ \theta $ respectively. As is  
well known this flow is completely integrable: 
\[ \{ p , t \} = 0 \,, \] 
and $ t = \xi ( \partial_\theta ) $ is called the Clairaut integral. 
Abstractly, $ \partial_\theta $ is the vector field generating  
the $ {\Bbb S}^1 $ action on $ X \setminus \{ m \}$.  
As in the case of compact simple surfaces of revolution  
(see \cite{Zel2},\cite{CdV},\cite{Bess}) we have as stronger  
statement: 
\begin{prop} 
\label{p:1} 
For $ ( X \setminus \{ m \} , g ) $ with the metric $ g $  
satisfying \eqref{eq:1.1}-\eqref{eq:1.3} there exist {\em global} 
action angle variables on $ T^* ( X \setminus \{ m \} ) $. 
\end{prop} 
Although it plays no part in the  
proof, this is worth presenting here as the global action variables 
are closely related to the asymptotics of the phase shifts. 
\begin{pf} 
The moment map  
\[ T^* ( X \setminus \{ m \} ) \ni  
( x, \xi ) \stackrel{P}{\longmapsto} \left(  
|\xi|_{g_x} , \xi ( \partial_\theta) \right) \in {\Bbb R}_+  
\times {\Bbb R} \] 
has the range given by the open set $ B = \{ ( b_1 , b_2 )  
:  
|b_2 | < b_1 \} $. For any $ ( b_1 , b_2 ) \in B $, $ P^{-1}  
( b_1 , b_2 ) $ consists of a $ {\Bbb R} \times {\Bbb S}^1 $ 
orbit of a single geodesic in $ T^* ( X \setminus \{ m \} $ 
(the $ {\Bbb R } $-action corresponds to the geodesic flow  
and the $ {\Bbb S}^1$-action to the $ \theta $-rotation).  
 
In the case of case of a {\em simple} surface of revolution, 
the global action variables, $ ( I_1 , I_2 ) $, are defined by  
\begin{equation} 
\label{eq:1.6} 
I_j  ( b ) = \frac{1}{2 \pi} \int_{ \gamma_j ( b) } \alpha \,, \ \  
\alpha = \xi \cdot dx \,,  
\end{equation} 
where $ ( \gamma_1 ( b) , \gamma_2 ( b ) ) $ is a global  
trivialization of the bundle $ H_1 ( P^{-1}  ( b) ,  
{\Bbb Z} ) $ of the homology groups along the the fibers of $ P$. 
When $ \gamma_1 ( b ) $ is chosen as the orbit of the $ {\Bbb S}^1  
$-action, then $ I_1 = \xi ( \partial_\theta ) $. In the case 
of non-compact surfaces discussed here the fibers are given by  
$ {\Bbb R} \times {\Bbb S}^1 $ and not by $ {\Bbb S} ^1 \times  
{\Bbb S}^1 $ (except for the degenerate case of the meridians, 
$ t = 0 $ where the fiber is $ ({\Bbb R} \setminus \{ 0 \}) \times 
{\Bbb S}^1 $  where $ 0 $ corresponds to the point $ m $).   
Consequently the integral for  $ I_2 $ given by \eqref{eq:1.6} 
diverges (for $ \gamma_1 $ we can still take the compact orbit of  
the $ {\Bbb S}^1 $-action). Hence we have to normalize the integral 
using the fact that the surface is asymptotic to a cylinder with  
$ a ( r ) \equiv 1 $. If we take $ \gamma_2 ( b)$ to correspond  
to a geodesic in $ P^{-1} ( b) $, then outside of the turning  
point $ \rho = 0 $ (or $ r = 0 $ for the degenerate case of the 
meridians) it can be parametrized by $ r $. Then $ \xi \cdot 
dx $ becomes $ \rho dr $ and we can put 
\begin{equation} 
\label{eq:1.7}  
I_2 ( b) = \frac{1}{\pi} \lim_{ R \rightarrow \infty }  
\int_0^R \left( b_1^2 - \frac{{b_2}^2} {a( r)^2 } \right)_+^{\frac12} 
dr - \int_0^R ( b_1^2 - b_2^2 )^{\frac12} dr \,,  
\end{equation} 
that is we normalize by subtracting the ``free" $ \rho dr $  
defined by $ \rho^2 + b_2^2 = b_1^2 $. From this we find the  
angle variables as in \cite{Zel2},\cite{CdV}.
\end{pf}

\section{Review of scattering theory} 
There are many ways of introducing the scattering matrix on  
a manifold of the type we consider. Since we only assume \eqref{eq:1.2}, 
$  X $ is not a $b$-manifold in the sense of Melrose -- see \cite{Mel}, 
\cite{Ch}. It is a manifold with a cusp metric at one end 
and a conic metric at the other -- see \cite{Mel}. We shall not however use 
this point of view here. Instead we will proceed more classically  
and we will define the scattering matrix using the wave operators -- 
see \cite{Ch} for an indication of the relation between the two  
approaches. As in the proof of Proposition \ref{p:1} we need a  
free reference problem 
\begin{equation} 
\label{eq:2.0} 
 X_0 \simeq {\Bbb R} \times {\Bbb S}^1 \,, \ \ g_0 = dr^2 + d\theta^2 \,. 
\end{equation} 
On $ X $ and $ X_0 $ we define the wave groups, $ U ( t ) $ and  
$ U_0 ( t ) $: 
\[ U ( t) \; :\; {\cal C}_{\rm{c}}^\infty ( X)  
\times {\cal C}_{\rm{c}}^\infty  ( X) \ni ( u _0 , u_1 )  
\longmapsto \left( u ( t) , D_t u ( t) \right) \] 
where  
\[ ( D_t^2 - \Delta_g ) u = 0 \, , \ \ u\rest_{t=0} = u_0 \,, \ \  
D_t u \rest_{t = 0 } = u_1 \,. \] 
The operators $ U ( t )$ extend as a unitary group to the energy space, 
$ {\cal H} ( X) $,  
obtained by taking the closure of $ {\cal C}_{\rm{c}} ^\infty ( X)  
\times {\cal C}^\infty_{\rm{c}} ( X ) $ with respect to the norm 
\[ \| ( u_0 , u_1 ) \|^2_E = \| \nabla u_0 \|_{L^2}^2 + \| u_1 \|_{L^2 }^2\,. 
\] 
The definition and properties of $ U_0 (t)$ are analogous.  
 
We then define the M{\o}ller wave operators 
\[ W_\pm \; : \; {\cal H} ( X_0 ) \longrightarrow {\cal H} ( X)  \,,  
\] 
by  
\[ W_\pm [ w ] = \lim_{t \rightarrow \pm \infty } U ( - t)  
\chi ( r ) U_0 ( t ) w \,, \ \ w \in {\cal H} ( X_0 ) \,, 
\]  
where $ \chi \in {\cal C}^\infty ( [ 0 , \infty ) ; [0,1] ) $,  
$ \chi ( r) \equiv 0 $ for $ r < 1 $ and $ \chi ( r) \equiv 1 $  
for $ r > 2 $, and where for $ r > 1 $ we used the obvious idenitification 
of the corresponding subsets of $ X $ and $ X_0 $. In the situation  
we consider the existence of $ W _\pm $ is quite straightforward and 
we choose the wave rather than the Schr\"odinger picture just for 
variety. The scattering operator is 
\begin{equation} 
\label{eq:scat} 
S \stackrel{\text{def}}{=} W_-^* W_+ \; : \; {\cal H} ( X_0 )  
\longrightarrow {\cal H} ( X_0 )  
\end{equation} 
and, as we will see below it is a unitary operator. 
 
When there is no pure point spectrum then the wave operators 
$ W_\pm $ are themselves unitary. In all situations 
they are partial isometries 
and $ W_\pm ^* = \lim_{t \rightarrow \pm \infty } U_0 ( - t )  
\chi ( r) U(t) $. The null space of $ W_\pm ^* $ is the span of  
the $ L^2 $ eigenfunctions of $ \Delta $. Under our assumptions  
there could only be finitely many such eigenfunctions. 
 
The wave operators have the intertwining properties: 
\[ W_\pm \left( \begin{array}{ll} \ 0 & I \\ 
\Delta_{g} & 0 \end{array} \right) 
= \left( \begin{array}{ll} \ 0 & I \\ 
\Delta_{g_0} & 0 \end{array} \right) W_{\pm} 
\ \Longrightarrow \ \left[ S ,  
\left( \begin{array}{ll} \ 0 & I \\ 
\Delta_{g_0} & 0 \end{array} \right) \right] = 0 \,. \] 
 
Since all operators commute with the generator of the $ {\Bbb 
S}^1 $ action, $ \partial_\theta $, we decompose $ S $ using the 
spectral decompositions of $ \Delta_{g_0 } $ and of $ \partial_\theta$. 
It is easy to check that  
\[ \left( \begin{array}{ll} \ 0 & I \\ 
\Delta_{g_0} & 0 \end{array} \right) = \int_{-\infty}^\infty  
\lambda d E_\lambda^0  \] 
where the Schwartz kernel of $ d E_\lambda^0 $ is given by  
\[ d E_\lambda^0 ( r, \theta; r' , \theta' )  
= \frac{\text{sgn}(\lambda)}{(2 \pi)^2 }  
\sum_{ n \in {\Bbb Z} } e^{ i n ( \theta -  
\theta' ) } \left( \begin{array}{ll} \; 0 & I \\ \lambda^2 & 0  
\end{array} \right) e^{ i \text{sgn} ( \lambda ) ( \lambda^2  
- n^2 )^{\frac12} ( r - r' ) } ( \lambda^2 - n^2)_+^{-\frac12}  
d \lambda \,. \] 
Because $ S$ commutes with the generator of the free propagator, 
$ U _0 ( t) $ we obtain the scattering matrix at fixed energy using 
the above spectral decomposition: 
\[ S = \int S ( \lambda ) d E_\lambda^0 \] 
and then the decomposition corresponding to the eigenvalues of  
$ \partial_\theta $: 
\[ S ( \lambda ) = \frac{1}{2 \pi } \sum_{ n \in {\Bbb Z } }  
S_n ( \lambda ) e^{ i n ( \theta - \theta' ) } \,. \] 
From the structure of $ d E_\lambda^0 $ it is clear that $ S_n (  
\lambda ) \equiv 0 $ for $ |n | > |\lambda| $. For $ |\lambda | 
= |n | $ we follow \cite{Ch} and put  
\[ S_n ( \lambda ) = \lim_{ \tau \rightarrow |\lambda|+} S_n (  
\text{sgn} (\lambda ) \tau ) \,. \] 
We also note that $ S_ n ( \lambda ) = S_{-n} ( \lambda ) $. 
 
For $ |n | \leq |\lambda | $, $ S_n ( \lambda) $ is a unitary  
operator, that is, it is given by multiplication by a complex number 
of unit length: 
\begin{equation} 
\label{eq:2.1} 
S_n ( \lambda  ) = e^  { 2 \pi i \delta_n ( \lambda ) }  
\end{equation} 
and the number $ \delta_n ( \lambda ) $ is the $n$th {\em phase 
shift} at energy $ \lambda^2 $. Another way to think about $ S (  
\lambda ) $ is as a diagonal unitary $ ( 2 n + 1 ) \times  
( 2 n + 1 ) $ matrix, where $ n = [|\lambda |]$: 
\[ S ( \lambda ) =  \left(  e^{2 \pi i \delta_{k} ( \lambda ) } 
\delta_{ k j } \right)_{ -n \leq k,j \leq n} \,. \]
A more ``down-to-earth'' definition, 
following the traditional way of introducing phase shifts in 
one dimensional scattering, is given through asymptotic
expansions in  \eqref{eq:2.4} below.

The uniform behaviour as $ k $ and $ \lambda $ go to infinity  
and $ k \ll \lambda $ is a well understood semi-classical  
problem. To describe it we separate variables in the eigen-equation 
of the Laplacian. We remark that this procedure can also provide 
direct proofs of the general scattering theoretical statements 
above. 
 
The Laplace operator is given by  
\[ \Delta_g = D_r^2  - i \frac{a ' ( r) }{a ( r) } D_r +  
\frac{1}{a(r)^2} D_\theta^2 \] 
and on the eigenspaces of $ D_\theta $ it acts as  
\begin{align} 
\label{eq:2.2} 
\begin{aligned} 
\Delta_n & = D_r^2 - - i \frac{a ' ( r) }{a ( r) } D_r +  
\frac{1}{a(r)^2} n^2 \\ 
& = a(r)^{-\frac12} \left( D_r^2 + \frac{n^2} { a( r) ^2 } 
- \frac{ 2 a '' ( r) a ( r) - ( a' ( r) ) ^2 }{ 4 a( r) ^2 }  
\right) a(r) ^{\frac12} \,. 
\end{aligned} 
\end{align} 
The reduced operator appearing in brackets in the second line 
above has a  self-adjoint realization  
on $ L^2 ( ( 0 , \infty)_r ) $ and for large $ \lambda $ it 
can be considered semi-classically: 
\begin{gather} 
\label{eq:2.3} 
\begin{gathered} 
\Delta_n - \lambda^2 = \lambda^2 a ( r) ^{-\frac12} P ( x, h )  
a ( r) ^{\frac12} \,, \ \ h = \frac{1}{|\lambda | } \,, \  x =  
\frac{n}\lambda \,, \\  
P ( x , h ) = ( h D_r )^2 + V ( r; x , h ) - 1 \,, \\ 
V ( r  ; x , h ) = \frac{x^2}{ a ( r )^2} - h^2  
\frac{ 2 a'' ( r) a ( r) - ( a' ( r) ) ^2 } { 4 a ( r) ^ 2 } 
\,, \ \ V_0 ( r ; x ) \stackrel{\text{def}}{=} V ( r; x , 0 ) \,. 
\end{gathered}
\end{gather} 
The principal symbol of $ P ( x, h ) $ is given by  $ p =  
\rho^2 + x^2/a ( r) ^2 - 1 $ and the natural range of $ x $ for 
which semi-classical methods are applicable is given by $  
0 < \epsilon < |x | < 1 - \epsilon $. In fact, since $ a (r  ) $  
is one at infinity, we approach zero energy when $ x^2 $ is close to  
$ 1 $. On the other hand when $ x \rightarrow 0 $ the characteristic 
variety of $ p $ has a singular limit -- see Fig.3.
A detailed analysis of the $ x \rightarrow 0 $ limit has to involve 
the lower order terms in $ V ( r; x , h ) $. In particular, 
miraculous cancellations in the expansions due to the interaction 
between the leading and lower order terms
occur when we have {\em product type} conic 
singularities since we can then use the theory of Bessel functions.  
The general situation is, at least to the authors, unclear at the moment. 
What is quite clear is that we have a uniform expansion in $ h/x $.

\begin{figure}[htb]
\hspace{0.0cm}
\epsfbox{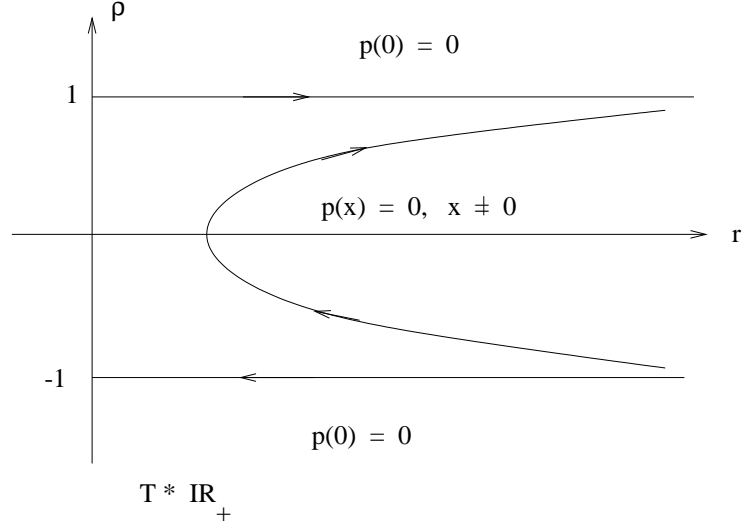}
\caption{The characteristic variety of $ P ( x, h ) $.}
\label{Fig.2}
\end{figure}

The phase shifts $ \delta_n ( \lambda ) $ are related to the 
semi-classical phase shifts of the operator $ P ( x  ,h ) $, 
$ \psi ( x, h ) $ which are defined by asymptotics of solutions: 
\begin{gather} 
\label{eq:2.4} 
\begin{gathered} 
P ( x, h ) u = 0 \,, \ \ 
u ( r ) = e^{\frac{i}h \sqrt{1 - x^2 } r } +  
e^{ \frac{i}{h } \psi( x, h ) }  e^{- \frac{i}h \sqrt{1 - x^2 } r } +  
{\cal O} \left( \frac{1}{r} \right) \,, \ \ r  
\rightarrow \infty\,, \\ 
\delta_n ( \lambda ) = \frac{|\lambda |}{2 \pi }  
 \psi \left( \frac{ n }{\lambda } , 
\frac{1}{|\lambda | } \right) \,. 
\end{gathered} 
\end{gather} 
We recall now the essentially standard asymptotic properties of 
$ \psi$ -- see \cite{Ol}, Chapter 6 and for a more microlocal 
discussion \cite{Ra}.
\begin{prop} 
\label{p:2} 
As $ h \rightarrow 0 $, $ \psi ( x , h ) $ defined by \eqref{eq:2.4} 
has an asymptotic expansion uniform in $ \epsilon < |x| < 1 -  
\epsilon $ for any fixed $ \epsilon > 0 $: 
\begin{equation} 
\label{eq:2.4'} 
\psi ( x , h ) \sim \psi ( x ) + h \frac\pi 2 + h^2 \psi_2 ( x) +  
\cdots  
\end{equation} 
where  
\begin{equation} 
\label{eq:2.4''} 
\psi ( x ) = \int_0^\infty \left( ( 1  - V_0 ( r , x)  )_+^{\frac12} 
- ( 1 - x^2 ) ^{\frac12} \right) dr \,.  
\end{equation} 
\end{prop} 
\stopthm

We remark that when we translate the asymptotics to the coordinates on $  
T ^* ( X \setminus \{ m\} ) $: $ x = t/\lambda $, $ \lambda^2 =  
\rho^2 + t^2/a(r)^2 $ we obtain the second action variable defined in  
the proof of Proposition \ref{p:1}. 
 
As mentioned in the introduction,  
we can describe this connection between the phase shifts and action variables
by saying that $S(\lambda)$ is a quantum map on $G(X,g)$, the space of geodesics.
We now digress to explain this statement in more detail.  For simplicity 
we will consider $ S $ at integral values of $ \lambda $ and denote them 
by $ N$. 

Since it is not needed in
the calculation of the limit pair correlation function we give a somewhat sketchy
discussion and refer to \cite{Zel1} for background on Toeplitz quantization.   See
also \cite{Sm0} for a related discussion from a physicist's point of view.

We can identify $G(X,g)$ with the set $S^*_{\rm{in}}(X_{r_0})$ of
incoming vectors at the parallel 
\[ X_{r_0} \stackrel{\rm{def}}{=}  X \cap \{ r = r_0 \} \,. \]
As in Proposition \ref{p:2}  we have to 
to delete $\epsilon$-neighborhoods of the singular set, given by 
$\{|t / \lambda |<  \epsilon\}$ where 
$ t = I_1 = 
\xi ( \partial_\theta ) $ is the first action variable  and 
$ \lambda^2 = \rho^2 + t^2 a(r)^{-1} $ is the energy.

We denote  the  deleted space of geodesics 
by $G_{\epsilon}(X,g)$ and identify it with 
the set $S^*_{\rm{in}, \epsilon}(X_{r_0})$ of
incoming vectors at $X_{r_0}$ 
with incoming angle satisfying $|\theta| > \epsilon$ and
$|\theta - \pi|> \epsilon.$  
If we consider the deleted space of geodesics as a phase space then, 
on the quantum level, it corresponds
to the sequence of truncated Hilbert spaces ${\cal H}_{N, \epsilon}$ spanned by the
eigenfunctions $\{e^{i n \theta}\}$ of the quantum action $\frac{1}{N}\hat{I}_1$ 
with $\epsilon < |\frac{n}{N}| < 1 - \epsilon$ where $\hat{I}_1 = -i 
\partial_{\theta}.$  Here,
${1}/{N}$ plays the role of the Planck constant and we restrict it to 
integral values.
Since ${\cal H}_{N, \epsilon}$ is invariant under $S(N)$ we may restrict the latter to
a unitary scattering matrix $S_{\epsilon}(N)$ on ${\cal H}_{N, \epsilon}$. 

 We now state the somewhat informal:

\begin{prop} 
\label{p:z}
The sequence $\{S_{\epsilon} (N)\}$ is a semiclassical quantum map over
$G_{\epsilon}(X,g)$ associated to the classical scattering map 
$$\beta: S^*_{\rm{in}, \epsilon}(X_{r_0}) 
\to S^*_{\rm{in}, \epsilon}(X_{r_0})$$
where $\beta(x, \xi)$ is obtained by following the geodesic $\gamma_{(x, \xi)}$ through
$(x, \xi)$ until it intersects $X_{r_0} $ for the last time and 
reflecting the outgoing
tangent vector inward. \end{prop}
\begin{pf}
From the explicit formula
\begin{equation} S_{\epsilon}(N) = (e^{2 \pi i \delta_k(N) } \delta_{kj})_{
\epsilon \leq |k/N|, |j/N| \leq (1 - \epsilon) },\;\;\;\;
\delta_k(N) = \frac{|N|}{2\pi} \psi \left(\frac{k}{N}, \frac{1}{N}
\right)  \end{equation}
we see that $S_{\epsilon}(N)$ is the exponential of $N$ times  
the  Hamiltonian 
\[ \widehat{H}_{N, \epsilon} = 
\chi_{\epsilon}\left(\frac{\widehat{I}_1}{N}
\right)
 \psi\left(\frac{\widehat{I}_1}{N}, \frac{1}{N}\right)\]
on ${\cal H}_{N, \epsilon}$  
where $\chi_{\epsilon}$ is a smooth cutoff function defining the truncated 
Hilbert space. 

The truncated phase space $G_{\epsilon}(X,g)$ is symplectically equivalent to a truncated $S^2$, equipped with its
standard area form, with neighbourhoods of the poles and of the equator deleted. Indeed,
the equivalence is defined by the identity map between 
 global action-angle charts on the surfaces.   This map intertwines the obvious
$S^1$ actions which rotate the spaces. 
 The quantization of this chart then defines a unitary equivalence
on the quantum level which intertwines the operators ${\partial}_{
\theta}$ on cylinder and sphere
(they can be considered as the angular momentum operators). 
The  equivalence is specified up to a choice
of $2N + 1$ phases by mapping the 
spherical harmonic of degree $N$ which transforms under  rotation by $\theta$
 on $S^2$ by $e^{i k \theta}$  to the exponential
$e^{i k \theta}$ with $k \in [-N, \dots, N]$.  The map is completely specified by
requiring that the spherical harmonic be real valued along $\theta = 0.$

Thus we may identify  $\widehat{H}_{N, \epsilon}$ with a Hamiltonian over 
the compact phase space $S^2.$ Since it is a function of the (Toeplitz) action operator $\widehat{I_1 }$ it
is necessarily a semiclassical Toeplitz operator of order zero with principal 
symbol
$\chi_{\epsilon}({I_1}/{E}) \psi({I_1}/
{E})$ on $S^2$.  
The semiclassical parameter $N$ is identified in the Toeplitz theory with a 
first order positive elliptic Toeplitz operator with eigenvalue $N$ in ${\cal H}_N$
-- see \cite{Zel1} and references given there. Hence 
$N \widehat{H}_{N, \epsilon}$ is a first order 
Toeplitz operator of real principal type.
As in the  essentially analogous case of pseudodifferential operators,
the exponential of a first order Toeplitz operator of real principal type is
a Fourier integral Toeplitz operator whose underlying classical map is the Hamilton
flow generated by $\psi$.   

We now wish to identify this map at time one with  the classical scattering
(or billiard ball) map on $S^*_{\rm{in}, \epsilon}(X_{r_0})$, 
that is,  we wish to prove that $\beta = \exp \Xi_{\psi}$ where $\Xi_{\psi}$ is
the Hamilton vector field of $\psi.$

Indeed, let us work in the
symplectic action-angle coordinates $(\theta, I_1)$ where $\theta$ is the angle along $X_{r_0}$.
The Hamilton flow of $\psi$ then takes the form
\begin{equation} \exp t \Xi_{\psi}(\theta, I_1) = (\theta + t \omega, I_1),\;\;\;\;
\omega = \partial_{I_1} \psi. \end{equation}
At time $t = 1$ the angle along the parallel $X_{r_0}$ changes by $\omega$.  We claim
that $\omega$ is also the change in angle along the incoming geodesic through $\theta \in X_{r_0}$ in the direction $I_1$ as it scatters in the bullet head before exiting again
along $X_{r_0}$.  

To see this, we use Proposition \ref{p:2} which shows that $ \psi $ is 
closely related to the second action variable:  in the notation of the  
proof of Proposition \ref{p:1} 
\[ I_2 ( b_1 , b_2 ) = b_1 \psi \left( \frac{b_2}  {b_1} \right) \,, \ \ 
I_1 ( b _1 , b_2 ) = b_2 \,. \]
Since $ b_1 = \lambda $ is preserved by the flow we can fix it at $ 
\lambda  = 1 $ and then
\begin{equation} \partial_{I_1} \psi (I_1) = 2  I_1 \int 
\frac{dr}{a(r)^2} \left(1 - \frac{I_1^2}{a(r)^2}
\right)_+^{-\frac{1}{2}}. \end{equation} 
On the other hand the equations of motion show that
\begin{equation} 
\label{eq:3.z}
\frac{d\theta}{dr}
 = \frac{ \dot \theta }{ \dot r }
= \frac{ 2 t a(r)^{-2} }{2 \rho } = 
 \frac{I_1}{a(r)^2} \left( 1 - \frac{I_1^2}{a(r)^2} \right)^{-\frac12} \,.
\end{equation}
It follows that $\omega(I_1)$ is twice 
the change in angle as the radial distance changes 
from $r_0$ to its minimum along the geodesic.

The piece of the geodesic lying in the bullet-head consists of two segments: the
initial segment beginning on $S_{r_0}$ and ending upon its tangential intersection
with the  parallel $X_{r_-(I_1)}$ closest to $m$, and the segment beginning at this
intersection and ending on $X_{r_0}.$  The change in $\theta$-angle along both segments
is the same, so that the total change in angle during the scattering is given by 
 the integral \eqref{eq:3.z} above.  This shows that $\beta$ and 
$\exp \Xi_{\psi}$ have precisely
the same formula in action-angle variables and completes the proof of the proposition.
\end{pf}

\section{Exponential sums} 
Following \cite{Zel1} we will reduce the study of \eqref{eq:pcm2} 
to a study of certain exponential sums. We first remark that 
because of symmetries of $ \delta_k ( \lambda ) $ we can study a  
slightly simpler expression 
\[ \tilde{\rho}^\epsilon_\lambda ( f) =  
\frac{1}{ ( 1 - 2 \epsilon) \lambda }  
\sum_{m \in {\Bbb Z} } \sum_{ \epsilon < j/\lambda, k/\lambda  
< 1 - \epsilon } f \left( ( 1 - \epsilon ) \lambda (  
\delta_j ( \lambda ) - \delta_k ( \lambda )  + m ) \right)  \]
as one easily checks that 
\[ \tilde\rho_\lambda^\epsilon ( f) = \frac{1}{2}  
\rho_\lambda^\epsilon \left( f \left( \frac12 \bullet \right)  
\right) \,. \] 
 
The reduction to exponential sums  
follows from an application of the Poisson summation formula in  
$ m $: 
\[ \tilde \rho_\lambda^\epsilon ( f) =  
\frac{1}{ [( 1 - 2 \epsilon ) \lambda]^2 } 
\sum_{ \ell \in {\Bbb Z} } \hat f \left( \frac{ 2 \pi \ell } 
{ ( 1 - 2 \epsilon ) \lambda } \right) \left| 
\sum_{ \epsilon < k/\lambda < 1 - \epsilon } e^{ i \ell 
\delta_k ( \lambda ) } \right|^2  \,. \]
From this we see that 
\begin{gather} 
\label{eq:4.4} 
\begin{gathered} 
\tilde \rho_\lambda^\epsilon ( f) =  
\hat f ( 0 ) + \int \hat f ( 2 \pi \xi ) d \xi  + o_{\lambda 
\rightarrow \infty} (1) + E^\epsilon_\lambda ( f ) \,, \\ 
E^\epsilon_\lambda ( f) \stackrel{\text{def}}{=} 
\frac{1}{ [( 1 - 2 \epsilon ) \lambda]^2 } 
\sum_{ \ell \in {\Bbb Z}\setminus \{ 0 \} } \hat f \left( \frac{ 2 \pi \ell } 
{ ( 1 - 2 \epsilon ) \lambda } \right)  
\sum_{ {\epsilon < k/\lambda, j /\lambda < 1 - \epsilon}\atop 
{ j \neq k } } e^{ i \ell 
(\delta_k ( \lambda ) - \delta_j ( \lambda ) ) }   \,.  
\end{gathered} 
\end{gather} 
Ideally, we would like to show that $ E^\epsilon_\lambda ( f)  
\rightarrow 0 $ as $ \lambda \rightarrow 0 $. That however seems 
very hard and for some surfaces is simply not true. As in  
\cite{Sa},\cite{RuSa},\cite{Zel1} we will instead consider families of 
surfaces and resulting families of scattering phase shifts: 
\[ ( \alpha_0 - \gamma , \alpha_0 + \gamma) \times 
( - \gamma , \gamma ) \ni ( \alpha , \beta )  \longmapsto 
\delta^{\alpha, \beta} _k ( \lambda ) \,. \] 
Replacing $ \delta_k ( \lambda ) $ by $ \delta_k^{\alpha, \beta} (  
\lambda ) $ in \eqref{eq:4.4} we define 
\begin{equation} 
\label{eq:4.5} 
 ( \alpha_0 - \gamma , \alpha_0 + \gamma ) \times 
( - \gamma , \gamma 
 ) \ni ( \alpha , \beta )  \longmapsto 
E_\lambda^\epsilon ( f; \alpha, \beta ) \,. 
\end{equation} 
To see the point of doing this we recall from \cite{Sa},\cite{RuSa} 
and \cite{Zel1} the following simple 
\begin{lem} 
\label{l:1} 
If for any $ f  \in {\cal S} ( {\Bbb R} ) $ 
\[ \int_{\alpha_0 - \gamma}^{\alpha_0 + \gamma} \int_{-\gamma} 
^\gamma | E_\lambda^\epsilon ( f; \alpha , \beta )|^2 d \alpha d \beta   
\leq  
C_{\epsilon, f} F ( \lambda ) \,, \] 
then for any sequence $ \{ \lambda_m \}_{m=0}^\infty $ such that 
\[ \sum_{m=0}^\infty F ( \lambda_m ) < \infty \] 
we have  
\[ E^\epsilon_{\lambda_m}  ( f ; \alpha, \beta ) \longrightarrow 0 \,, \ 
m \longrightarrow \infty \,, \ \  
\forall \;  f \in {\cal S}  ( {\Bbb R} ) \] 
almost everywhere in $ ( \alpha , \beta ) \in ( \alpha_0  
- \gamma , \alpha_0 + \gamma ) \times ( - \gamma , \gamma)  $. 
\end{lem} 
\stopthm
 
When $ \delta^{\alpha, \beta}_k ( \lambda ) $ have a somewhat  
idealized form, the crucial estimate comes from \cite{Zel1}, Theorem 5.1.1 
where it is loosely based on the {\em Vinogradov method}.
Since we will need a further 
development of these estimates we present a slightly modified proof. 
\begin{prop} 
\label{p:3} 
If in \eqref{eq:4.4} and \eqref{eq:4.5}  
\[ \delta_k^{\alpha, \beta} ( \lambda ) =  
\alpha {k} 
+ \beta {\lambda} \Phi \left( \frac{k}{\lambda } \right) \,, \ \ 
\Phi \in {\cal C}^\infty ( ( 0 , 1 ) )\,, \ \  
| \Phi'' \rest_{( \epsilon , 1 - \epsilon ) } | \geq C_\epsilon  
> 0 \,, \]
then for any $ f \in {\cal S} (  
{\Bbb R} ) $
\[ \int_{-1}^{1} \int_{-1}^{1} | E^\epsilon_\lambda ( f ;  
\alpha , \beta )|^2 d\alpha d \beta = {\cal O}_{f, \epsilon } 
\left( \frac{\log ^3 \lambda }{\lambda } \right) \,,  
\ \ \lambda \longrightarrow \infty \,.  
\] 
\end{prop} 
\begin{pf} 
Let $ \rho_\delta  \in {\cal C}^\infty  ( {\Bbb R} ) $ have the  
following properties 
\[ \rho_\delta (t) \geq \bbbone_{[-1,1]} ( t) \,, \ \  
\text{supp} \; \hat \rho_\delta \subset ( - \delta , \delta ) \,. \] 
The estimate of the lemma will clearly follow from  
\begin{equation} 
\label{eq:4.5'} 
\int_{\Bbb R} \int_{\Bbb R}  \rho _\delta ( \alpha )  
\rho_\delta ( \beta) \left| E_\lambda ^\epsilon ( f ;  
\alpha , \beta ) \right|^2 d \alpha d \beta = {\cal O} \left( 
\frac{ \log^3  \lambda} {\lambda} \right)  \,.  
\end{equation} 
Using the representation of $ E_\lambda^\epsilon $, \eqref{eq:4.4}, 
the left hand side of \eqref{eq:4.5'} can be rewritten as  
\begin{gather}
\label{eq:4.rew} 
\begin{gathered}
 \frac{1}{\lambda^{4} } \sum_{\ell_1 \neq 0 } \sum_{ \ell_2 \neq 0 } 
g \left( \frac{\ell_1}{\lambda } \right) 
\overline{g \left( \frac{\ell_2}{\lambda } \right) }  
 \sum_{ {\epsilon < k_1/\lambda , j_1/\lambda < 1 - \epsilon } 
\atop{ k_1 \neq j_1 } }  
\sum_{ {\epsilon < k_2/\lambda , j_2/\lambda < 1 - \epsilon } 
\atop{ k_2 \neq j_2 } }  \\
 \hat \rho_\delta \left( \ell_1 ( j_1 - k_1 ) - \ell_2 ( j_2 - k_ 2 )  
\right)  
\hat \rho_\delta \left( \lambda \left( \ell_1  
\left( \Phi \left( \frac{j_1}{ \lambda } \right) -  
\Phi \left( \frac{k_1 }{\lambda } \right) \right)  -  
\ell_2  \left( \Phi \left( \frac{j_2}{ \lambda } \right) -  
\Phi \left( \frac{k_2 }{\lambda } \right) \right) \right)  \right)  \,, 
\end{gathered} 
\end{gather} 
where we dropped the overall factor of $ ( 1 - 2 \epsilon)^{-2} $  
and put $ g ( \xi ) \stackrel{\text{def}}{=} 
 f ( 2 \pi ( 1 - 2 \epsilon )^{-1} \xi ) $. From now on we will 
drop the parameter $ \epsilon $ altogether: we can for instance extend
$ \Phi $ as a strictly convex or concave function to $ [0,1] $ adding
additional positive terms to the sums which are being estimated or 
we can shift and rescale the variables.
 
The support condition on $ \hat \rho_\delta $ implies that 
\begin{gather} 
\label{eq:4.e} 
\begin{gathered}
\ell_1 ( j_1 - k_1 ) - \ell_2 ( j_2 - k_2 ) = 0 \\ 
\left|   
\ell_1  
\left( \Phi \left( \frac{j_1}{ \lambda } \right) -  
\Phi \left( \frac{k_1 }{\lambda } \right) \right) -  
\ell_2  \left( \Phi \left( \frac{j_2}{ \lambda } \right) -  
\Phi \left( \frac{k_2 }{\lambda } \right) \right)  \right| 
\leq \frac{\delta}{\lambda} \,.  
\end{gathered} 
\end{gather} 
To understand the second expression we apply the mean value theorem  
twice to the difference of $ \Phi $'s. For that we make a simple
observation: 
if $  \phi'' $ has a fixed sign on 
$ [a - \epsilon , b + \epsilon ]  $  then if
$ (m+h)/2, ( m-h)/2    \in [a , b ] $  
\begin{gather*}
\phi \left( \frac{m + h}{2}
 \right) - \phi  \left( 
\frac{ m -h }{2} \right) = h \phi' ( \Xi ( m , h ) )  \,, \\
\frac{1}{2} \frac{ \min |\phi'' |}{\max 
|\phi'' | }  \leq  
\frac{\partial{\Xi}}
{\partial m } ( m , h ) 
\leq \frac{1}{2} \frac{ \max |\phi'' |}{\min |\phi'' | } \,. 
\end{gather*} 
In our case we put $ \phi = \Phi ( \bullet / \lambda ) $ and 
$ h_i = j_i - k_i \neq 0 $ and $ m_i = j_i + k_i $. Then 
with 
\[ \xi_{\lambda , h } ( m ) \stackrel{\text{def}}{=}  \Xi (m, h ) \]
we have
\begin{equation}
\label{eq:4.calc'}
1/C < \partial_m \xi_{ \lambda, h } ( m )  < C \,, 
\end{equation}
and \eqref{eq:4.e} implies
\begin{gather} 
\label{eq:4.e''} 
\begin{gathered} 
\ell_1 h_1 = \ell_2 h_2 \,, \ \ 0 < |h_i | \leq \lambda  \\ 
\left| \ell_1 h_1 \left( m_2 - \xi_{ \lambda, h_2}^{-1} 
\left( \xi_{ \lambda , h_1} ( m_1  ) \right) \right) 
\right|  \leq C \delta \lambda \,, \ \ 0 \leq m_i \leq 2 \lambda \,, 
\end{gathered} 
\end{gather} 
where we can invert $ \xi_{\lambda, h_2} $ in view of \eqref{eq:4.calc'}.

Thus we want study 
the sets of six integers $ ( h_i , m_i , l _i ) $, 
$ i = 1,2$, satisfying  \eqref{eq:4.e''}.
We first note that, say, $ h_2 $ is determined by $ h_1, \ell_1,  
\ell_2 $. Then we see from the second inequality in \eqref{eq:4.e''}
and from \eqref{eq:4.calc'} that for fixed $ (h_1 , h_2, m_1,  \ell_1) $
there are 
\[  {\cal O} ( 1)  \max \left( 1, \frac{ \lambda}{| \ell_1 h_1| } \right)  \ \ 
\text{$m_2$'s satisfying \eqref{eq:4.e''}} .\]

When $ \lambda > |\ell_1 h_1 | $ the contribution to \eqref{eq:4.rew}
is estimated by 
\begin{align*}
\frac{1}{\lambda^4} \sum_ {1 \leq m_1 \leq 2 \lambda}
\sum_{ {-\lambda \leq h_1 \leq \lambda}\atop{ h_1 \neq 0 } }
\sum_{ \ell_1 \neq 0 } \sum_{ \ell_2 \neq 0 } 
\frac{ \lambda} {| \ell_1 h_1|} 
\left|  g \left( \frac{\ell_1}{\lambda } \right)\right| 
\left| g \left( \frac{\ell_2}{\lambda } \right) \right| & \leq 
C \delta \frac{\log \lambda} {\lambda} \int_{ |\xi| >  \lambda } 
| g ( \xi ) | \frac{ d\xi }{ |\xi| } \int | g ( \xi ) | d \xi \\ 
& \leq C_g \delta \frac{\log^2 \lambda }{\lambda}  \,.
\end{align*} 

When $ \lambda \leq |\ell_1 h_1| $ then the number of $ m_2 $'s is 
uniformly bounded for each choice of the other variables. We want to 
count the 
triples $ ( h_1 , h_2 , \ell_1 ) $ satisfying the first equation  
of \eqref{eq:4.e''} as a function of $ \ell_2 $ and $ \lambda $. 
Let $ F ( \lambda , \ell_2 ) $ denote that number. 
If $ d (n )$ denotes the number of divisors of $ n \neq 0 $ then
\[ F ( \lambda , \ell_2 ) \leq 8 \sum_{ 0 < h_2 \leq    \lambda}
d ( h_2 |\ell_2  |) \]
since $ \ell_1 h_1 = \ell_2 h_2 $ and each factorization into a
product has to be counted twice since $ \ell_1 $ and $ h_1 $ can 
be interchanged.
Then 
\begin{align*} G ( \lambda , N ) & \stackrel{\text{def}}{=} 
 \sum_{0 \neq |\ell_2 | \leq N } F({ \lambda , \ell_2 })
  \leq  8 \sum_{ \ell_2 \neq 0 } \sum_{ 0 < h_2 \leq \lambda }
d ( |\ell_2 | h_2 ) \\
& \leq C \sum_{ 1 \leq n \leq N \lambda } d ( n)^2  
\leq C' \lambda N ( \log \lambda + \log N )^3 \,, \end{align*} 
by a theorem of Ramanujan - see \cite{HW}, Sect.18.2 and references 
given there.
Hence the part of \eqref{eq:4.rew} corresponding to the bounded number
of $ m_2 $'s is bounded by 
\begin{align*}
 C \lambda^{-4 } \sum_{ 1 \leq m_1 \leq 2 \lambda } 
\max |g |   \sum_{ \ell_2 \neq 0 }  
F( \lambda , \ell_2 )  \left| g \left( \frac{\ell_2}{\lambda } \right) 
\right| 
& \leq C'_g \lambda^{-3} \int \left( G( \lambda , \lambda |\xi| )   + 1 
\right) 
|g' ( \xi )| d\xi  \\ 
& \leq C_g \frac{\log^3 \lambda}{\lambda}  \,,\end{align*} 
where we used summation by parts and then approximation by the
Riemann integral.
This completes the proof of the proposition. 
\end{pf}

We now recall from Proposition \ref{p:2} that for surfaces we  
consider we have  
\begin{equation} 
\label{eq:4.7} \delta_{k} ( \lambda ) = \lambda \psi \left( \frac{k}{\lambda} 
\right) + \frac14 + \frac{1}{\lambda} \psi_2 \left( \frac{k}{\lambda}, 
\frac{1}{\lambda} \right) \,, \ \ \epsilon < \frac{k}{\lambda}  
< 1 - \epsilon \,, \ \ \lambda \rightarrow \infty \,, \end{equation} 
where for $ \epsilon < x < 1 - \epsilon $, $ \partial_x^k \psi_2  
= {\cal O}_{k, \epsilon} ( 1) $.  
 
When the family of surfaces depends on two parameters, $  ( \alpha,  
\beta ) $, so that \eqref{eq:1.1}-\eqref{eq:1.3} hold uniformly, then 
we also have \eqref{eq:4.7} uniformly with respect to the parameters. 
Hence to apply Proposition \ref{p:3} to our case we need to estimate 
the contribution of the error terms coming from $ \psi_2^{\alpha, \beta} $. 
That is given in  
\begin{prop} 
\label{p:4} 
Let $ \delta^{\alpha, \beta}_k ( \lambda ) $ be given by  
\eqref{eq:4.7} uniformly in $ ( \alpha , \beta ) \in ( \alpha_0  
- \gamma , \alpha_0 + \gamma ) \times ( - \gamma , \gamma ) $ 
with  
\[ \psi^{\alpha , \beta}  ( x  ) = \alpha x + \beta \Phi ( x) \,, \ \ 
\Phi \in {\cal C}^\infty ( ( 0 , 1 ) )\,, \ \  
| \Phi'' \rest_{( \epsilon , 1 - \epsilon ) } | \geq C_\epsilon  
> 0 \,. \]
Then 
\[  \int_{\alpha_0-\gamma}^{\alpha_0 +\gamma} \int_{-\gamma} 
^\gamma | E_\lambda^\epsilon ( f; \alpha , \beta )|^2 d \alpha d \beta   
 =  {\cal O}_{f , \epsilon } \left( \frac{ \log^3 \lambda }
{\lambda  }  \right) \,.  \]
\end{prop} 
\begin{pf} 
We observe that $ \psi^{\alpha, \beta } $ is defined for all  
$ ( \alpha, \beta ) $ and that we can extend $ \delta_k ^{\alpha,  
\beta } $ to all $ ( \alpha, \beta ) $ by smoothly cutting off the lower 
order terms for $ ( \alpha, \beta ) \in {\Bbb R} \times  
{\Bbb R} \setminus ( - \gamma - \epsilon , \gamma +  
\epsilon  ) \times (  \alpha_0 - \gamma - \epsilon , \alpha_0 +  
\gamma + \epsilon ) $. 
Using the inequality $ |x|^2 \leq 2 |y|^2 + 2 |x-y|^2 $ and  
Proposition \ref{p:3} we see that 
we need to estimate 
\begin{gather*}  \int \int \left| \frac{1}{\lambda^2 } 
\sum_{\ell\neq 0 } g \left( \frac{\ell}{\lambda} \right)   
\sum_{{ \epsilon < k/\lambda ,  j/\lambda < 1 - \epsilon } 
\atop { k \neq j } } \left( e^{ i \ell ( \tilde \delta_k^{ 
\alpha, \beta } ( \lambda )  - \tilde \delta_j ^{ \alpha, \beta} 
( \lambda ) ) } -  
e^{ i \ell (  \delta_k^{ 
\alpha, \beta } ( \lambda )  -  \delta_j ^{ \alpha, \beta} 
( \lambda ) ) } \right)  \right|^2  \rho_\delta ( \alpha )  
\rho_\delta ( \beta ) d \alpha d \beta \,, \\
\tilde \delta^{\alpha, \beta  } _k 
( \lambda ) \stackrel{\text{def}}{=} 
\lambda \psi^{\alpha, \beta} ( k / \lambda ) \,, 
\end{gather*} 
where we use the notation of the proof of Proposition \ref{p:3}. 
We now introduce 
\[ \tau ( z) \stackrel{\text{def}}{=}  2i \sin \frac{z}{2} \exp  
\left( -  i\frac{z}{2} \right) 
\,, \ \ \ e^{ i x } - e^{ i y } = e^{ i x } \tau ( x-y ) \,, \ \ 
\frac{\tau ( z) } z \in {\cal C}^\infty \,,  
\] 
and put 
\begin{align*}
 \psi_{ \ell, j , k , \lambda } ( \alpha , \beta ) & =  
\frac{\lambda}{\ell} \tau \left( \ell \left(  
\delta_k^{\alpha, \beta} (\lambda) - \delta_j 
^{\alpha, \beta}  ( \lambda ) - \tilde \delta_k^{\alpha, \beta} 
 ( \lambda)  
- \tilde \delta_j^{\alpha, \beta} 
 ( \lambda ) \right) \right) \\
& = \frac{\lambda}{\ell} \tau \left( \frac{\ell}{\lambda} 
\left( \psi^{\alpha, \beta} _2 \left( \frac{k}{\lambda} , 
\frac1{\lambda} \right)
-  \psi^{\alpha, \beta} _2 \left( \frac{j}{\lambda} , \frac1{\lambda} \right)
\right) \right) 
\,. \end{align*} 
From Proposition \ref{p:2} and the obvious properties of $ \tau $ 
we see that $ \psi_{ \ell, j , k , \lambda } $ is $ {\cal C}^\infty $  
and that it satisfies the following estimates 
\begin{equation} 
\label{eq:4.8'} 
\left| \partial^{p_\alpha}_\alpha \partial^{p_\beta}_\beta  
\psi_{\ell, j , k , \lambda } ( \alpha, \beta ) \right|  
\leq C_p \left( 1 + \left| \frac{\ell}{\lambda} \right| \right)^{  
p_\alpha + p_\beta } \,. 
\end{equation} 
Hence we have to look at 
\[  \int \int \left| \frac{1}{\lambda^2 } 
\sum_{\ell\neq 0 }  
\left( \frac{\ell}{\lambda} \right)   
g \left( \frac{\ell}{\lambda} \right)   
\sum_{{ \epsilon < k/\lambda ,  j/\lambda < 1 - \epsilon } 
\atop { k \neq j } }  
\psi_{ \ell, j , k , \lambda } ( \alpha , \beta )  
 e^{ i \ell ( \tilde \delta_k^{ 
\alpha, \beta } ( \lambda )  - \tilde \delta_j ^{ \alpha, \beta} 
( \lambda ) ) }   \right|^2  \rho_\delta ( \alpha )  
\rho_\delta ( \beta ) d \alpha d \beta \,. \] 
 
We would like to proceed as in the proof of Proposition \ref{p:3} 
but now taking of Fourier transforms has to be replaced 
by integration by parts. Using the analysis of the differences of  
$ \tilde \delta_k^{\alpha, \beta } $ presented there we need to  
estimate 
\begin{equation}
\label{eq:4.gath}
 \frac{1}{\lambda^2} \sum_{\ell_1 \neq 0 } \sum_{\ell_2 \neq 0 }  
\left| \frac{\ell_1}{\lambda}   
g \left(  \frac{\ell_1}{\lambda}  \right) \right| 
\left| \frac{\ell_2}{\lambda}   
g \left(  \frac{\ell_2}{\lambda}  \right) \right|
 \langle \ell_1/\lambda \rangle^2 
\langle \ell_2/\lambda \rangle^2 I ( \lambda , \ell_1 
, \ell_2 )  \,, \end{equation}
where  
\[ I  ( \lambda , \ell_1 , \ell_2 )  
= \sum_{ m_1 = 1}^{ 2 [\lambda] }  
 \sum_{ m_2 = 1}^{ 2 [\lambda] }  
\sum_{ {h_1 = - [\lambda ]} \atop {h_1 \neq 0   }}^{[\lambda] }
\sum_{ { h_2 = - [\lambda ]}  \atop { h_2 \neq 0 }} ^{[\lambda] }
\langle \ell_1 h_1 - \ell_2 h_2 \rangle ^{-2}  
 \langle \lambda^{-1} (   \ell_2 h_2 \xi_{\lambda , h_2} ( m_2)  -   
\ell_1 h_1  \xi_{ \lambda , 
h_1 } ( m_1 )  ) \rangle^{-2} 
\,, \] 
and where, as is usual, we write $ \langle x \rangle  =  
( 1 + x^2 )^{\frac12} $. We note that we can absorb the terms 
$ \langle \ell_1 / \lambda \rangle^2 $ and $ \langle \ell_2 / 
\lambda \rangle^2 $ into the $g$ terms.

We first observe that
\[ \sum _{  m \in {\Bbb Z} } \langle A - B m \rangle^{-2} = 
{\cal O} ( 1) \max \left( 1 , \frac{1}{ B } \right) \,,\]
uniformly in  $ A \in {\Bbb R} $. In fact, for $ |B| \leq 1  $  this
follows from the comparison with the integral using the Euler-MacLaurin
formula
\[ \sum_{-\infty }^{\infty } f(n ) = \int_{-\infty}^\infty f( x ) dx +
{\cal O} \left( \int_{-\infty}^\infty |f'' ( x) | d x  \right) \,, \]
and for $ |B| > 1 $  we can write  the sum as $ \sum_{k} ( 1 + 
B^2 ( A/B -[A/B] -k )^2 ) ^{-1} = {\cal O}(1) $.

Using this and 
\eqref{eq:4.calc'} we see that
\[ \sum_{ m \geq 1 } \langle A - B \xi_{\lambda,h }(m)  \rangle^{-2} 
\leq C \sum_{m \geq 1} \langle B m - B \xi_{ \lambda , h } ^{-1}
( A/ B) \rangle^{-2} \leq C \max \left( 1 , \frac1{B} \right) \,.\]
Hence, uniformly in $ \ell_2 , h_2 $
\begin{equation}
\label{eq:4.new} 
\sum_{m_1} 
 \langle \lambda^{-1} (   \ell_2 h_2 \xi_{\lambda , h_2} ( m_2)  -   
\ell_1 h_1  \xi_{ \lambda , 
h_1 } ( m_1 )  ) \rangle^{-2} 
 = {\cal O} ( 1) \max \left( 
1 , \frac{ \lambda } { |h_1 \ell_1| } \right) \,.
\end{equation} 
 
Proceeding as in the proof of Proposition \ref{p:3} we introduce 
 $  \widetilde F 
( \lambda ,  \ell_2, p  ) $ as the number of $ ( \ell_1, 
h_1, h_2 ) $ satisfying  $ \ell_1 h_1 = \ell_2 h_2 + p $.
We now have 
\[ \widetilde F ( \lambda ,  \ell_2, p  )  \leq 
4 \sum_{{ 0 \neq |h_2| \leq \lambda } \atop 
{ \ell_1 h_2 + p \neq 0 } } d ( |\ell_2 h_2 + p | ) \,,\]
and 
\begin{align*}  \widetilde G ( \lambda , N ) & \stackrel{\text{def}}{=}
\sum_{ p = - \infty}^\infty 
\sum_{ |\ell_2 | \leq N } \widetilde F({ \lambda , \ell_2 }, p )
\langle p \rangle^{-2} \\
& \leq  C_1 \sum_{ p = - \infty }^\infty \sum_{{1 \leq |n| \leq 
N \lambda} \atop { n + p \neq 0 } } d (|n|) d ( |n + p |) \langle 
p \rangle^{-2}  \\
& \leq C_2 \left( \sum_{  n =1}^{ N \lambda } d(n)^2 \right)^{\frac12}
\sum_{p=0}^\infty  \; \langle p \rangle^{-2}
\left( \sum_{ m=1}^{ N \lambda + p }  d ( m )^2 
 \right)^{\frac12} \\
& \leq  C_3  \left( \lambda N ( \log \lambda + \log N )^3 \right)^{\frac12}
\sum_{p = 0 }^\infty  \; \langle p \rangle^{-2} \left( 
(\lambda N + p )  \log^3 (\lambda  N + p ) 
\right)^{\frac12} \\
& \leq C  \lambda N ( \log \lambda + \log N )^3 \,. \end{align*} 
Using \eqref{eq:4.new} we can estimate 
\eqref{eq:4.gath} by:
\[ C   \max (1 +  |\xi|) | g ( \xi ) |  \lambda^{-3} \int 
\left(
\widetilde   G( \lambda , \lambda |\xi| )  + 1 \right) 
|\partial _\xi ( \xi g) ( \xi )| d\xi  
\leq C_g \frac{\log^3 \lambda}{\lambda}  \,,\] 
 completing the proof of the proposition. 
\end{pf} 

\medskip
\noindent
{\bf Remark.} As was emphasized by the referee, Propositions 
\ref{p:3} and \ref{p:4} are essentially sharp. The ``diagonal''
solutions $ j_1 = j_2 $, $ k_1 = k_2 $, $ l_1 = l_2 $ give a 
lower bound $ 1/ \lambda $. Hence, no essential improvement of 
the Theorem in Sect.1 is possible by this method.
 
\section{Construction of the family of surfaces} 
To prove the main theorem we need to construct a family of 
surfaces, $ {\cal G} = \{ g^{\alpha, \beta} \}$, for which in the  
expansion of the phase shifts \eqref{eq:4.7},  
\begin{equation} 
\label{eq:5.1} 
 \psi^{\alpha, \beta} ( x ) 
= \alpha x + \beta \Phi ( x ) \,, \ \  
|\Phi''\rest_{( \epsilon , 1 - \epsilon ) } | > C_\epsilon >  0\,, \ \  
( \alpha, \beta ) \in ( \alpha_0 - \gamma , \alpha_0 + \gamma ) 
\times ( - \gamma , \gamma ) \,, 
\end{equation} 
so that we can apply Propositions \ref{p:3} and \ref{p:4}. 
Recalling Proposition \ref{p:2}, this means that we want to  
find $ a ^{\alpha, \beta } $ satisfying \eqref{eq:1.1}-\eqref{eq:1.3}, 
and such that  
\begin{equation}
\label{eq:5.2} 
 \psi^{\alpha, \beta} ( x ) 
=
\frac{1}{ \pi }  
 \int_0^\infty \left( \left( 1 - \frac{x^2}{a^{\alpha,\beta} 
 (r)^2} \right)_+^{\frac12}  
- ( 1 - x^2)^{\frac12} \right) dr = \alpha  x + \beta \Phi ( x) \,,  
\end{equation} 
with $ \Phi $ convex or concave.  
 
We will now skip the indices  
$ \alpha $ and $ \beta $. If we write  
\[ W ( r) \stackrel{\text{def}}{=} \frac{1}{a( r) ^2 }- 1  \,, \] 
and 
\begin{equation} 
\label{eq:5.3} 
 \phi ( x ) \stackrel{\text{def}}{=}  \int_0^\infty \left(  
\left( 1 - \frac{x^2 }{ W ( r)} \right)^ 
{\frac12}_+ -  1 \right) d r \,,  
\end{equation} 
then  
\[ \psi ( x) = \frac{1}{ \pi } 
\sqrt{1 - x^2} \phi \left( \frac{x}{\sqrt{1-x^2}} \right)\,,\] 
and we might study the simpler function $ \phi $ instead. 
From the assumptions on $  a $,  $ W $ is monotonically decreasing and  
$ r^2 W ( r) $ is smooth and non-zero at $ r = 0 $. Hence there exists 
a smooth monotonically increasing function, $ y ( r) $, such that 
\[ W(r) = \frac1{ y ( r )^2 } \,. \] 
Since we can write $ r $ as a function of $ y $ we define 
\[ F ( y ) \stackrel{\text{def}}{=} \frac{dr}{dy } ( y ) \,.\] 
That way we can express $ \phi ( x ) $ as a linear transform of $F$: 
\begin{equation} 
\label{eq:5.4} 
\phi ( x  ) = x I ( F) ( x) \,, \ \ 
I ( F ) ( x ) = \int_0^\infty \left( \left( 1 - \frac1{y^2} \right)_+^ 
{\frac12} - 1 \right) F ( x y ) dy \,. \end{equation} 
From this we immediately get a {\em linear model} correspoding to  
$ F ( x ) \equiv t > 0 $: 
\[ a ( r ) = \frac{r^2}{ t^2 + r^2 } \ \ \ \Longrightarrow  \ \ \ \psi ( x)  
= \frac{\sqrt{1-x^2}}{\pi} \phi \left( \frac{x}{\sqrt{1-x^2} } \right) 
= -\frac{1}{2} t x \,, \] 
since 
\[ \int_0^\infty \left( \left( 1 - y^{-2} \right)^{\frac12}_+ - 1  
\right) dy  = - \pi/2 \,. \] 

\vspace{0.2cm}
\noindent
{\bf Remark.} The surfaces defined using the linear model do {\em not}
have uniform pair correlations measures. In that case we can compute
the leading contribution to $ E_\epsilon^\lambda ( f ) $ directly.
To apply Propositions \ref{p:3} and \ref{p:4} we need to have the 
linear term in $ \psi $ and that forces the singularity at $ 0$ 
for our surfaces: only one value of $ \alpha $ corresponds to a smooth
surface.
\vspace{0.2cm}

We want to introduce the convex or concave term in $ \psi $ by  
perturbing the case $ F \equiv \text{const} $. For that let us  first 
establish some simple properties of the transform $ F  
\mapsto I ( F ) $. We denote by $ {\cal C}^\infty_{\rm{b}}  $ 
smooth functions with bounded derivatives and by $ S^k_{\rm{phg}} $ 
spaces of poly-homogeneous (classical) symbols. 
 
\begin{lem} 
\label{l:2} 
For $ g \in {\cal C}^\infty_{\rm{b}} ( [0, \infty )) $ 
\[ I ( g)  ( x ) \in {\cal C}^\infty_{\rm{b}} ( [0, \infty ))  
+ x \log x {\cal C}^\infty_{\rm{b}} ( [0, \infty ))  \,. \] 
When $ g \in S^{-2}_{\rm{phg}} ( [0 , \infty )) $ then 
\[ I ( g)  \rest_{[1, \infty ) } \in S^{-1}_{\rm{phg}} ( [1,  
\infty ) ) \, , \ \  
I ( g ) ( x ) \sim \left( - \int f( y ) dy \right) \frac{1}{x}  
+ \sum_{k=2}^\infty \frac{g_k  }{x^k} \,, \ \ x \longrightarrow 
\infty \,. \] 
\end{lem} 
\begin{pf} 
To prove the first part of the lemma we write 
\[ I (g) ( x) =  \int_0^C \left( ( 1 - y^{-2} )_+^{\frac12}  
- 1 \right) g ( xy ) dy  
+ \int_C^\infty \left( ( 1 - y^{-2} )^{\frac12}  
- 1 \right) g ( xy ) dy  \,, \ \ C > 1 \,, \] 
where the first term on the right hand side is clearly in  
$ {\cal C}^\infty_{\rm{b}} ( [0 , \infty )) $. In the  
second term, the integrand can be rewritten as  
\[ \frac{1}{ y^2 } \left( - \frac12 + \frac{1}{8} \frac{1}{y^2}  
-\cdots \right) g ( xy ) dy \,. \] 
Thus we are concerned with integrals of the form  
\[ \int_C^\infty \frac{1}{y^k} g ( x y ) dy =  
x^{k-1} \int_1^\infty Y^{-k} g ( Y ) d Y  +  
x^{k-1} \int_{Cx}^1 Y^{-k} g ( Y ) d Y  \,.\] 
where the first term is smooth and uniformly bounded in $ k $. 
To study the second term 
we write  
\[ g ( Y ) = g_0 + g_1 Y + \cdots g_{l-1} Y^{l-1} + \tilde g_l ( Y )Y ^l \] 
which 
gives 
\[ \frac{1}{k-1} C^{-k+1} g_0 + \frac1
{k-2} C^{-k+2} g_1 x +  
\cdots +  \frac{C^{-k+l}}{k-l} g_{l-1}x^{l-1}  + x^{ l} F_{k,l, C} ( x) \,, \] 
for $ k > l  $ and 
\[ \frac{1}{k-1} C^{-k+1} g_0 + \frac1
{k-2} C^{-k+2} g_1 x +  
\cdots + g_{k-1}x^{k-1} \log x  +  x^{k-1} G_{k,l,C} ( x)   \,, \] 
for finitely many $ k \leq l $. Since we check that
\begin{gather*}
G_{k,l,C} ( x) \in {\cal C}^\infty ( [ 0 , 1/C )) \,, \ \ k \leq l \,, \\
 |F_{ k,l,C} ( x) | \leq G_l ( C^{ -k + l + 1 } + |x|^{ k-l -1 } ) \,, \ \ 
k > l \,, 
\end{gather*}
we can sum up the contributions from different $ k$'s ($ C > 1$, $ |x| 
\ll 1 $, and we use the uniform convergence of $ ( 1 - z)^{\frac12} $). 
Thus for every $ l $ we obtain 
\[ I ( g) ( x) = h_{1,l} ( x )  + x \log x h_{2, l } ( x )  + {\cal 
O} ( x^l ) \,, \ \ h_{1,l}, h_{2,l} \in {\cal C}^\infty \,,\]   
and consequently $ I ( g) \in {\cal C}^\infty_{\rm{b}} (  
[0, \infty )) + x \log x {\cal C}^\infty_{\rm{b}}  ( [0, \infty )) $. 
 
The second part of the lemma is even more clear. If for large $ Y $,  
$ g ( Y ) \sim \sum_{k=2}^\infty {G_k}{Y^{-k}} $, then
then  
\[ \int_0^\infty  \left( ( 1 - y^{-2} )_+^{\frac12} - 1 \right)  
g ( x y )\; \sim_{x \rightarrow \infty } \;  
 - \frac{1}{x}   \int g ( Y ) dY +  
\sum_{k=2}^\infty \frac{1}{x^k} \left( \int_1^\infty \left( 1 -  
\frac{1}{y^2} \right) ^{\frac12} \frac {f_k}{y^k} dy \right) \,.\] 
\end{pf} 
 
The lemma shows that we cannot expect smoothness of  $ \psi ( x) $ 
at the end points $ x=0,1$ but that the function is very well  
behaved in the interior, as in any case is implicit in Proposition  
\ref{p:2}. 
 
Having discussed the general properties of the transform $ I$ 
we now state a straightforward  
 
\begin{lem} 
\label{l:3} 
For $ \Phi ( x ) = x I ( f) ( x/ \sqrt{1 - x^2 } ) $  
and $ 0 < x ( z) = z/\sqrt{1 + z^2} < 1 $   
we have 
\[ \Phi'' ( x ( z) )  = ( 1 + z^2)^{\frac32} 
\left[ \int_0^\infty \left( ( 1 - y^{-2} )_+^{\frac12} - 1  
\right) \left( ( 2 + 3 z^2 ) y f' ( y z ) + z y^2 ( 1 + z^2 )  
f'' ( z y ) \right) dy \right] \,.  \]
\end{lem} 
\stopthm
 
Guided by the two lemmas we can easily construct a family of 
surfaces for which $ \psi $ has the needed properties. We want to 
find $ f \in {\cal C}^\infty_{\rm{b}} ( {\Bbb R} ) $ such that
for $ \beta $ small enough and $ \alpha $ close to $ \alpha _0 $,
$ F^{\alpha, \beta } ( y  ) = \alpha  + \beta f ( y ) > 0 $,  
and so that $ a ( r ) $ obtained from inverting the process described
above 
has the properties \eqref{eq:1.1}-\eqref{eq:1.3}. 
This is easily achieved by demanding that $ f  $ is a symbol of
order $ -2 $ on $ [ 0 , \infty ) $.

We also want 
$ \Phi'' ( x ) $ described in Lemma \ref{l:3} to have a fixed sign.  
From the formula we see that that  $ \Phi $ is concave if
\begin{equation}
\label{eq:5.c}
 a f'(  y  )  + y f'' ( y ) > 0 \, , \ \ y > 0 \,, \ \ a= 2, 3 \,.
\end{equation}
In fact, the integrand has the same sign as 
\[ ( 2 + 3 z^2 )  f' ( Y ) + ( 1 + z^2 ) Y f '' ( Y ) \rest_{Y=yz} \]    
and that is positive for any $ z $ if \eqref{eq:5.c} holds.

We can summarize this discussion in 
\begin{prop}
\label{p:5}
For $ \alpha $ in a neighbourhood of a fixed $ \alpha_0 < 0 $ and 
for $ \beta $ small enough, let
$ a^{\alpha, \beta }_f  (r) $ be obtained from 
$ f \in S^{-2}_{\rm{phg}} ([0, \infty )) $ by the following procedure:
\begin{gather*}
   F^{\alpha , \beta }  ( y )   = - 2 \alpha - \beta f ( y ) \,, \\
\frac{dy^{\alpha, \beta} }{dr} = (F^{\alpha , \beta } ( y^{\alpha, 
\beta } ( r) ) ^{-1} \,, \ y^{\alpha, \beta } ( 0 ) = 0 \,, \ 
a^{\alpha, \beta} ( r ) = y^{\alpha, 
\beta } ( r) ( 1 +  y^{\alpha, 
\beta } (r)^2)^{-\frac12} \,. \end{gather*} 
Then $ a^{\alpha, \beta}_f $ has the properties \eqref{eq:1.1}-\eqref{eq:1.4} 
and for the set of two parameter families of surfaces 
\[ {\cal G} = \{ ( X, g^{\alpha, \beta} _f ) \; : \;
g_f^{\alpha, \beta } = dr^2 + a^{\alpha, \beta} _f ( r) d \theta ^2 \,, \ \ 
f \in S^{-2}_{\rm{phg}} ( [0, \infty )) \ \text{{\em satisfies} 
\eqref{eq:5.c}},
\  | \alpha - \alpha_0 | < \delta_f \,, \ |\beta | < \epsilon_f \,  
 \}\]
the leading part of the phase shifts depends linearly on $ \alpha$ and 
$ \beta $ and \eqref{eq:5.1} holds. 
\end{prop}

Combined with Propositions \ref{p:2} and \ref{p:4} 
this provides an infinite dimensional family of perturbations of the linear
model each giving a two parameter family of surfaces for which the Theorem  
of Sect.1 holds. Perhaps the simplest example is obtained by putting 
\[  F^{\alpha , \beta }  ( y )   = - 2 \alpha - \beta \frac{1}{1 + \rho y^2
} \,, \ \ 0 < \rho \leq \frac23 \,. \]

\vspace{0.15cm}

\noindent
{\em Proof of the Main Theorem.} 
Proposition \ref{p:5} guarantees that the leading parts of the expansions
of the phase shifts of $( X, g^{\alpha, \beta} ) $ satisfy the assumptions
of Proposition \ref{p:4}. 
Let $ \rho^{\epsilon, \alpha, \beta}_\lambda $ denote the pair correlation
measure for $ ( X, g^{\alpha, \beta} ) $ given by \eqref{eq:pcm2}. 
Recalling \eqref{eq:4.4} and the discussion 
preceding it we see that all for $ f \in {\cal S} ( {\Bbb R} )$ 
\[ \rho^{\epsilon, \alpha, \beta}_\lambda ( f ) 
= \hat f (0 ) + f ( 0 ) + o_{\lambda
\rightarrow \infty } ( 1 ) + E_\lambda^\epsilon ( f; \alpha, \beta )  \,. \]
Proposition \ref{p:4} now shows that the assumptions of Lemma \ref{l:1}
are satisfied with $ F ( \lambda )  = \log^3 \lambda / \lambda $ and
that lemma gives the statement of the Main Theorem.
\stopthm

\vspace{0.25cm}
{\sc 
\noindent
The Johns Hopkins University \\
\vspace{0.2cm}

\noindent
University of California, Berkeley\\
and\\
University of Toronto}
 
\end{document}